\documentclass[preprint,12pt]{elsarticle}

\usepackage{amssymb, amsmath}
\usepackage{amsthm}

\graphicspath{{img/}}
\usepackage{subcaption}

\usepackage[utf8]{inputenc}
\usepackage[T1]{fontenc}

\usepackage[separate-uncertainty = true]{siunitx}
\DeclareSIUnit\gauss{G}
\newcommand{\EeV}{\exa\electronvolt}

\usepackage{booktabs}

\usepackage{microtype}
\usepackage{lmodern}

\journal{Astroparticle Physics}

\begin{document}

\begin{frontmatter}



\title{Determination of the absolute energy scale of extensive air showers via radio emission: systematic uncertainty of underlying first-principle calculations}


\author[buw]{Marvin Gottowik\corref{cor1}}
\ead{gottowik@uni-wuppertal.de}

\author[rwtha,cal]{Christian Glaser}
\ead{christian.glaser@uci.edu}

\author[kit,bru]{Tim Huege}
\ead{tim.huege@kit.edu}

\author[buw]{Julian Rautenberg}
\ead{julian.rautenberg@uni-wuppertal.de}

\cortext[cor1]{Corresponding author}

\address[buw]{Bergische Universität Wuppertal, Gaußstraße 20, 42119 Wuppertal, Germany}
\address[rwtha]{RWTH Aachen University, III. Physikalisches Institut A, Aachen, Germany}
\address[cal]{Department of Physics and Astronomy, University of California, Irvine, USA}
\address[kit]{Karlsruher Institut f\"ur Technologie, Institut f\"ur Kernphysik, Postfach 3640, 76021 Karlsruhe, Germany}
\address[bru]{Vrije Universiteit Brussel, Astrophysical Institute, Pleinlaan 2, 1050 Brussels, Belgium}

\begin{abstract}
Recently, the energy determination of extensive air showers using radio emission has been shown to
be both precise and accurate.
In particular, radio detection offers the opportunity for an independent measurement of the absolute
energy scale of cosmic rays, since the radiation energy (the energy radiated in the form of radio signals)
can be predicted using first-principle calculations involving no free parameters, and the
measurement of radio waves is not subject to any significant absorption or scattering in
the atmosphere.
To quantify the uncertainty associated with such an approach, we collate the various contributions to the uncertainty, and we verify the consistency of radiation-energy calculations from microscopic simulation
codes by comparing Monte Carlo simulations made with the two codes CoREAS and ZHAireS.
We compare a large set of simulations with different primary energies and shower directions and
observe differences in the radiation energy prediction for the 30--80~MHz band of \SI{5.2}{\percent}.
This corresponds to an uncertainty of \SI{2.6}{\percent} for the determination of the absolute cosmic-ray energy scale.
Our result has general validity and can be built upon directly by experimental efforts for the calibration of the cosmic-ray energy scale on the basis of radio emission measurements.
\end{abstract}


\end{frontmatter}


\section{Introduction}

Ultra-high energy cosmic rays (UHECRs) impinging on the Earth's atmosphere induce huge cascades of secondary particles, so-called extensive air showers. Established techniques for their detection are the measurement of the particles that reach the ground as well as the observation of Cherenkov and fluorscence light emitted by the showers \citep{Bluemer2009293}. An important observable for most analyses of high-energy cosmic rays is their energy. It is particularly challenging to pin down the absolute scale of this energy measurement.

Particle detectors need to rely on hadronic interaction models that still exhibit large uncertainties at the highest energies. To avoid this, in particular the fluorescence technique is used to determine the absolute energy scale. Telescopes measure the fluorescence light emitted by air showers. This is proportional to the calorimetric shower energy and allows for an accurate determination of the absolute energy scale with a systematic uncertainty of as good as 14\% in case of the Pierre Auger Observatory \cite{EnergyScaleICRC2013}. However, fluorescence light detection is only possible at sites with very good atmospheric conditions, and precise quantification of scattering and absorption of fluorescence light under changing atmospheric conditions requires extensive atmospheric monitoring efforts.

Another method for the measurement of cosmic rays is the detection of broadband radio emission from air showers \cite{Huege:2016veh}. The radio technique combines a duty cycle close to 100\% with a precise measurement of the cosmic-ray energy as well as a good sensitivity to the mass of the primary cosmic ray \cite{Huege:2016veh,SchroederReview}. Furthermore, radio detection has the potential for an accurate determination of the absolute cosmic-ray energy scale \cite{Aab:2016eeq, TunkaRexCrossCalibration, KrauseICRC2017}. This is possible because there is no relevant absorption of radio waves in the atmosphere, and because the absolute strength of the radio emission can be calculated from first principles given the development of the electromagnetic cascade of an air shower. These first-principle calculations are thus a crucial ingredient in the determination of the energy scale via radio emission.

In addition to the absolute scale predicted by these calculations, also their systematic uncertainties have to be carefully evaluated. Due to the fundamental nature of the calculations, their intrinsic uncertainty can be estimated independent of a specific experiment. At a time where measurement uncertainties decreased below the 10\%-level \cite{AERAOctocopterJINST}, and consequently even minor uncertainties in the calculations become relevant, we perform a thorough determination of the systematic uncertainty of first-principle calculations of air-shower radio emission. Our result is directly relevant to and usable in all future endeavors to calibrate the absolute energy scales of cosmic-ray detectors via the radio technique.

Two mechanisms contribute to the coherent radio emission \cite{Huege:2016veh}. The dominant geomagnetic emission induced
by time-varying transverse currents in the Earth’s magnetic field $\vec{B}$ is polarized in the direction of 
the Lorentz force ($\vec{v} \times \vec{B}$) with shower direction denoted by $\vec{v}$.
The time-varying negative charge excess in the shower front is due to the knock-out of electrons
from air molecules and annihilation of positrons in the shower front and gives rise to radiation
polarized radially towards the shower core.

In simulations, the emitted radiation can be calculated directly from the movement of the shower
particles using classical electrodynamics \cite{Huege:2016veh} without the
need to explicitly model the two emission mechanisms.
Hence, the calculations do not involve any free parameters. However, the Monte Carlo simulation codes use approximations in the simulation of the shower development for computational feasibility, and the numerical implementation of classical electrodynamics might add additional uncertainties. In this work, we estimate the resulting uncertainties by a detailed comparison of the two independent simulation codes CoREAS \cite{coreas} and ZHAireS \cite{zhaires}.

The radiation energy -- the energy emitted by the air shower in the form of radio waves -- serves as a universal estimator of the cosmic-ray energy that is independent of the experimental setup and location of a radio detector \cite{Aab:2016eeq} and is already successfully exploited by the Pierre Auger Collaboration \cite{Aab:2015vta}. The radiation energy originates solely from the
well-understood electromagnetic part of the air shower and its determination results in a theoretical energy resolution of 3\% \cite{Glaser:2016qso} which is well below current experimental uncertainties. 
Hence, in the following we use the radiation energy to quantify the accuracy of the calculation of the radio emission in the comparison between CoREAS and ZHAireS.

\section{Systematic Uncertainties of the Prediction of the Radiation Energy}
To put our analysis into context, we first recapitulate the main steps needed to determine the cosmic-ray energy from a radio measurement and the systematic uncertainties currently associated with these. As outlined above, the experimental quantity of interest is the radiation energy. So far, only the Pierre Auger Collaboration has measured the radiation energy and reported a systematic uncertainty of 28\% \cite{Aab:2015vta,Aab:2016eeq}. This translates to an uncertainty on the absolute cosmic-ray energy of 14\%, as the radiation energy scales quadratically with the cosmic-ray energy. The experimental uncertainty is mainly driven by the uncertainty of the antenna response pattern which was recently reported to have reached a level of 9\% (with respect to the cosmic-ray energy) \cite{AERAOctocopterJINST, KrauseICRC2017}.

The measured radiation energy can then be translated to the energy in the electromagnetic cascade of an air shower (the electromagnetic shower energy) using predictions from air-shower simulation codes \cite{Glaser:2016qso}. Other air-shower components such as neutrinos and muons do not contribute to the radio emission. Hence, the final step is the conversion from the electromagnetic shower energy to the cosmic-ray energy. A theoretical prediction of the \emph{invisible} energy connecting the two is difficult as it depends on hadronic interaction models at high energies where their uncertainties are large. However, the Pierre Auger Collaboration has demonstrated that the invisibe energy fraction can be directly determined from data by combining measurements of the fluorescence telescopes and particle detectors, with a systematic uncertainty of 3\% \cite{ThePierreAuger:2013eja}.

The purpose of this work is a thorough determination of the systematic uncertainty in the calculation of the radiation energy from the electromagnetic shower energy.
Several aspects of this uncertainty have already been studied elsewhere.
The current state of these studies is summarized in Tab.~\ref{tab:systuncertainty} and will be shortly summarized in the following.

\paragraph{Hadronic interaction models} As only the relation between radiation energy and electromagnetic shower energy is of interest, the theoretical prediction is basically independent of the hadronic interaction model. This was studied in reference \cite{Glaser:2016qso} by exchanging the high-energy model QGSJETII-04 with EPOS LHC and the low-energy model FLUKA with UrQMD. The simulated radiation energies per electromagnetic energy agreed within 0.3\%.

\paragraph{Approximations in the air-shower simulation} As a full simulation of all shower particles is not feasible at high cosmic-ray energies due to the large number of shower particles, only a representative sub-sample of particles is tracked.
The effect of this approximation was studied in reference \cite{Glaser:2016qso}, and a suitable \emph{thinning level} that is sufficiently accurate was determined.
A finer simulation (lower thinning level, lower maximum particle weights) then leads to the same amount of radiation energy.

Another approximation is that only shower particles above a certain low-energy threshold are followed. In reference \cite{Glaser:2016qso} it was found that this approximation influences the radiation energy by less than 1\% for the low energy thresholds used in this work.

Here, we continue the important investigation of the influence of approximations in the air-shower simulations and examine the last unstudied approximation that is currently known: the impact of a finite step-length in the simulation code, discussed in section \ref{sec:stepsize}. 

\paragraph{State of the atmosphere} A thorough evaluation of the effects of the state of the atmosphere on the radiation energy depends on the exact location of an experiment and needs to be performed by the authors of the respective experiment. However, the general impact of the atmospheric conditions was determined in reference \cite{Glaser:2016qso}. The relevant quantity for radio emission is the refractivity. If the refractivity at sea level (which is then scaled with density to higher altitudes) is varied by $\pm$10\%, which is already larger than the yearly fluctuations at current experimental sites, the radiation energy changes by $\pm$3\% only. 

\paragraph{Underlying first-principle calculations}
The remaining uncertainty that has not been studied previously and is needed to obtain the complete systematic uncertainty, is the intrinsic uncertainty of the underlying first-principle calculations. This remaining aspect is studied thoroughly in this work by comparing predictions of the two independent codes CoREAS and ZHAires for configurations and input parameters set as similar as possible.

While comparisons of the two simulation codes have been made in the past, and an agreement of the predicted radiation energies at the 5\% level has been reported in reference \cite{Aab:2016eeq}, none of the comparisons performed so far have been systematic.
The value reported in reference \cite{Aab:2016eeq}, e.g., was only determined for one particular cosmic-ray energy and one particular event geometry.
Furthermore, differences in the underlying calculations such as different treatments of the refractive index profile of the atmosphere, were not addressed in the comparison.
Here, for the first time, we present a systematic study of the uncertainties in the first-principle calculations of the radiation energy of extensive air showers.

We find an agreement between CoREAS and ZHAireS within 5.2\% in the calculation of the radiation energy.
Correspondingly, the uncertainty in the determination of the absolute energy scale amounts to 2.6\%.
The uncertainty of the first-principle calculations thus turns out to dominate the overall theoretical uncertainty, cf.\ table \ref{tab:systuncertainty}.
Nevertheless, the theoretical uncertainty remains lower than the experimental uncertainty of $\approx 9$\% achieved to date. The details of the comparison are reported in the following.

\begin{table}
\newcommand\tw{0.5cm}
\newcommand\tww{0.2cm}
\centering
\begin{tabular}{lc}
 \toprule
  {source of uncertainty} & \vtop{\hbox{\strut uncertainty of}\hbox{\strut radiation energy}}\\
 \midrule
  hadronic interaction models \cite{Glaser:2016qso} & 0.3\% \\
  \rule{0pt}{4ex}approximations in the air-shower simulation &    \\
  \hspace*{\tww}particle thinning \cite{Glaser:2016qso} & $<0.3\%$ \\
  \hspace*{\tww}energy thresholds of shower particles \cite{Glaser:2016qso} & $<1\%$ \\
  \rule{0pt}{4ex}state of the atmosphere \cite{Glaser:2016qso} & $<3\%$ \\
  \textbf{underlying first principle calculations} & \textbf{<5.2\%} \\
 \bottomrule
\end{tabular}
\caption{Systematic uncertainties of the calculation of the radiation energy from the electromagnetic shower energy.}
\label{tab:systuncertainty}
\end{table}

\section{Simulation Setup}
For the simulation of the air showers, the two different codes CORSIKA 7.4100 and Aires 2.8.4a 
are used. The extensions CoREAS and ZHAireS are enabled for the computation of the radio emission.
Both codes calculate the radio emission from pure electrodynamics applied to each particle in
the simulation, but use different formalisms. In CoREAS the ``endpoint formalism''
\cite{2011PhRvE..84e6602J} is used, whereas in ZHAireS both time-domain and frequency-domain implementations of the ``ZHS 
algorithm'' \cite{Zas:1991jv, 2010PhRvD..81l3009A} are available. In this work, the time-domain
implementation is used. A previous comparison of the electric field strengths predicted by the two formalisms
for particle showers in a dense target showed agreement \cite{SLACT510-PRL} to within better than 5\%.
The Earth's curvature is taken into account by both codes.

An accurate model of the atmosphere is needed for a proper simulation of the radio emission. For
both codes we use a 5-layer model, namely Linsley's parametrization of the US standard atmosphere
as detailed in the CORSIKA user's guide \cite{corsika}.
For the scaling of the air refractivity $n-1$ to higher altitudes CoREAS adopts $n-1$ as
proportional to the air density, while ZHAireS uses a simple exponential scaling. To eliminate
effects of the refractivity model in the comparison, the CoREAS code has been modified to use a
ZHAireS-like simple exponential scaling for the refractivity.

To model the high energy interactions we consistently use SIBYLL 2.1 in both cases as this is the
only model available in both air shower simulation codes. Although SIBYLL 2.1 is outdated and produces different results 
than the updated version SIBYLL 2.3 we can nevertheless use this model to compare the underlying 
computation of the radio emission. However, the results presented here should not be used to infer an absolute relation between the radiation energy and the electromagnetic shower energy.

We simulate 1000 proton- and 1000 iron-induced air showers with each code. The primary energies
follow a uniform distribution of the logarithmic energy between \SI{e17}{\eV} and \SI{e19}{\eV}.
The azimuth angle is distributed uniformly between \SI{0}{\degree} and \SI{360}{\degree} and the
zenith angle uniformly between \SI{0}{\degree} and \SI{75}{\degree}. The geomagnetic field is set
to an inclination of \SI{-35.7}{\degree} with a field strength of \SI{0.243}{\gauss}. The refractivity
at sea level is set to $n-1 = 2.92 \times 10^{-4}$. These settings correspond to the 
environmental conditions of the Engineering Radio Array \cite{Schroder:2016imf} of the Pierre Auger Observatory \cite{ThePierreAugerCollaboration2015172}.

\subsection{Calculation of the Radiation Energy}
All simulations used in this study are band-passed with a digital filter to the 30--80~MHz band, which is the typical band used
in cosmic-ray radio detectors today. To ensure an identical bandwith after the filtering the CoREAS traces are zero-padded to match the length of the ZHAireS traces.

The computing time of the simulations increases almost linearly with the number of simulated
antennas. Thus, it is unfeasible to simulate signals for large numbers of antennas and sample the full
two-dimensional emission pattern for a big set of simulated air showers. 
To extract the charge-excess and geomagnetic components individually one can simulate nearly the same 
showers with and without a magnetic field \cite{Alvarez-Muniz:2014wna}.
Here, we follow the method developed in \cite{Glaser:2016qso}: It is sufficient to simulate only 
antennas on the  $\vec{v} \times (\vec{v} \times \vec{B})$ axis, as here the polarization of the 
charge excess and the geomagnetic components decouple. 

Assuming that the geomagnetic and the charge excess components are in phase and that 
their individual lateral distribution functions are radially symmetric we can calculate the total 
radiation energy as an integral of the energy fluence $f$ as
\begin{equation}
  \label{eq:Erd}
   E_{\text{rad}} = 2\pi \int_0^\infty \mathrm{d}r \, r \; f(r).
\end{equation}

In the following, 30 antennas are placed on the $\vec{v} \times (\vec{v} \times \vec{B})$ axis in
each simulation such that the complete radio-emission footprint is covered. We follow the prescription of \cite{Glaser:2016qso} to adjust the antenna spacing which ensures an adequate sampling 
of the radio signal distribution, such that the uncertainty of the numerical integration can be neglected.

This one-dimensional approach results in a slight and well-understood overestimation in the radiation energy compared to a two-dimensional integration over the complete radio footprint \cite{Glaser:2016qso}. To exclude any potential difference between CoREAS and ZHAireS resulting from the one-dimensional approximation, we resimulate a subset of 250 air showers on a full two-dimensional grid and compare with the one-dimensinal simulation. An overestimation of \SI{3.36}{\percent} (\SI{3.12}{\percent}) is found for CoREAS (ZHAireS) when using the one-dimensional calculation. To correct for this bias, $E_{\text{rad}}$ will be reduced accordingly in the following analysis.

\subsection{Influence of Technical Parameters of the Simulation Codes}
\label{sec:stepsize}
To make the simulations of both codes comparable, we set technical parameters to the same values or ensure that they have no influence on the radiation energy. In particular, parameters that control approximations in the MC code are chosen such that the radiation energy is not impacted. For CoREAS, the influence of several technical parameters has already been studied in \cite{Glaser:2016qso}. This included in particular the low-energy thresholds of the particles in the air-shower simulation, the thinning level, and the choice of hadronic interaction model. As the ZHAireS codes handles some approximations differently, we repeat some of these checks here.

In particular, an individual study is made to exclude any influence of the different thinning algorithms used in the simulations.
In CORSIKA, a thinning level of \num{e-5} with optimal weight limitation \cite{Kobal2001} is found to be sufficient. In Aires, the thinning is controlled by two parameters, the thinning level and a statistical weight factor. A thinning level of \num{e-5} with a statistical weight factor of \num{0.06} is found to be sufficient.
We also reaffirmed that the choice of the low-energy hadronic model has no influence on the radiation energy in CoREAS. No significant differences were found between the two tested models FLUKA \cite{fluka} and UrQMD \cite{0954-3899-25-9-308}.
Hence, UrQMD is used for convenience.

CoREAS applies the endpoint formalism \cite{2011PhRvE..84e6602J} for the calculation of the radio emission. This formalism becomes numerically unstable very near the Cherenkov angle, where CoREAS thus falls back to a treatment equivalent to the ZHS formalism \cite{Zas:1991jv}. We verified that the threshold at which this fallback is made in CoREAS does not influence the radiation energy. As the ZHS algorithm relies on the Fraunhofer approximation which requires that the simulation step is small with respect to the wavelength and distance from the observer to the shower particle, we also verified that the step sizes used in the ZHS-fallback calculation are indeed chosen small enough. 

Another approximation that influences the radiation energy and which has not been considered in previous work is the step size in the Monte Carlo simulation of the electromagnetic cascade. In CoREAS, the current default values were tuned to achieve a good compromise between computational speed and accuracy for the demands of particle detectors and fluorescence telescopes. For an accurate calculation of the radio emission we find that a finer simulation of the electromagnetic cascades is required. In contrast, the ZHAireS code already uses step sizes tailored explicitly to accomodate radio simulations, but unfortunately does not allow for a modification of the step sizes via an external parameter. Furthermore, there is no simple way to consistently modify the code itself \cite{MunizTuerosPrivateComms}. We therefore focus on the CoREAS code here.

\begin{figure}
\centering
\includegraphics[width=.8\linewidth]{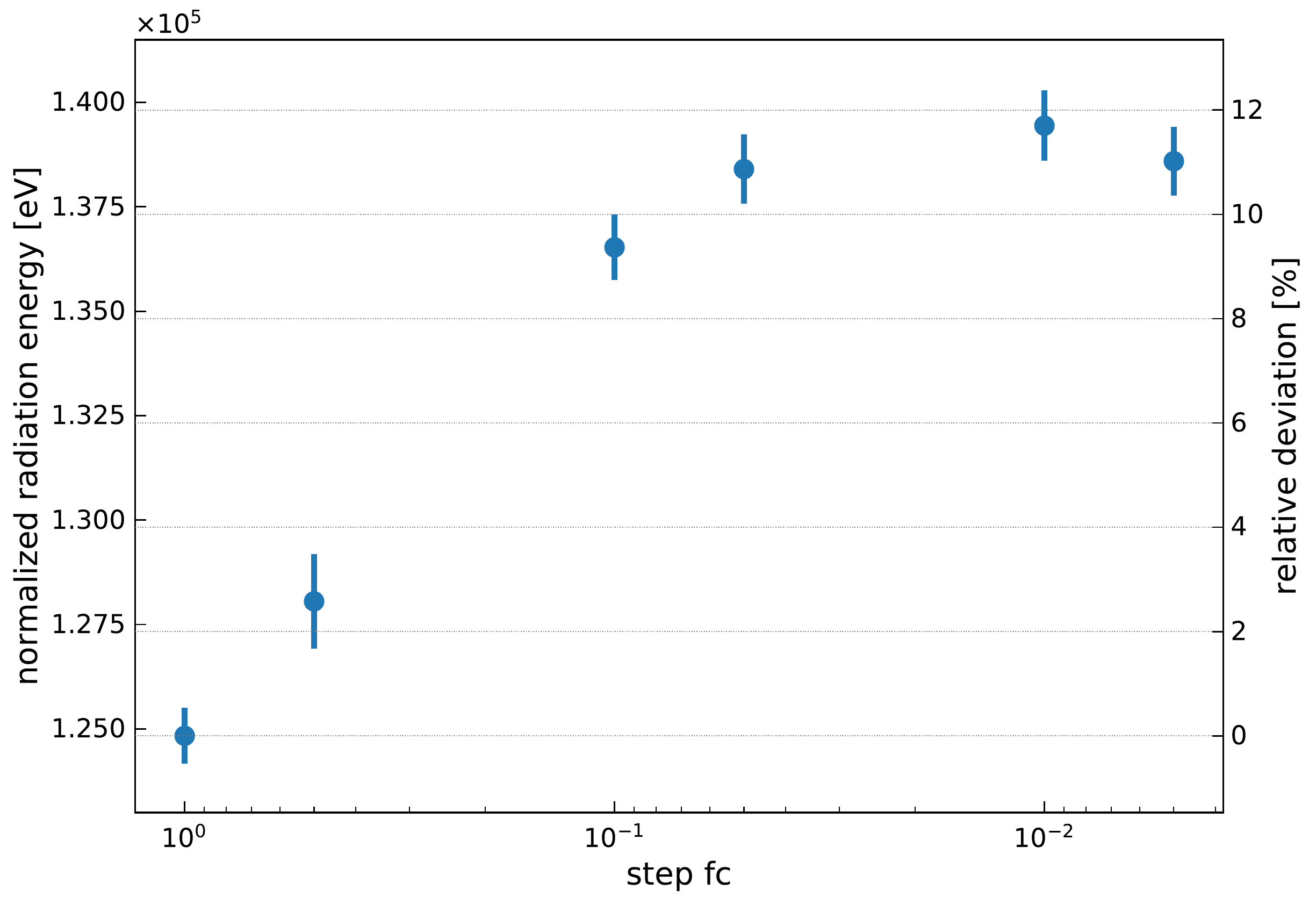}
\caption{Influence of the STEPFC parameter of CORSIKA on the radiation energy for an \SI{e17}{\eV}
air shower simulated with CoREAS. Shown is the absolute value and the relative deviation from the default setting. The errorbar shows the uncertainty of the mean radiation energy of 40 air shower simulations. }
\label{fig:coreas_stepfc}
\end{figure}

In CORSIKA, the ``STEPFC'' parameter allows us to modify the electron multiple scattering length factor in the underlying EGS4 simulation from the default setting. By reducing this factor, we achieve a finer simulation of the particle cascade. 
We analyze the impact on the radiation energy by studying 
proton-initiated air showers with an energy of \SI{e17}{\eV} and a zenith angle of \SI{50}{\degree} coming from 
the south. For each different setting of the parameter, we simulate 40 air showers to minimize the impact of shower-to-shower fluctuations.
We observe that the radiation energy increases and converges towards a level of 11\% above the value obtained with default settings when we decrease the step sizes up to a factor of 20 (i.e., set STEPFC = 0.05). For even smaller step sizes (we simulated air showers with up to a factor of 200 smaller step sizes), the radiation energy remains at the same level (cf.~Fig.~\ref{fig:coreas_stepfc}). In a separate analysis we confirmed that this increase in radiation energy is a global factor which is independent of the geometry and the energy of the air shower. 

To judge the consistency of the step sizes in CORSIKA and in ZHAireS, we studied the average step sizes adopted in both codes as a function of altitude. We find that the step sizes in ZHAireS are slightly larger than in CoREAS with STEPFC = 0.05. At an altitude of \SIrange{4}{5}{km} the average step size amounts to 0.34~m (0.49~m) in CORSIKA (ZHAireS). At an altitude of \SIrange{11}{12}{km} the average step size corresponds to 0.68~m(0.67~m). For even higher altitudes, the step size in ZHaireS becomes smaller than in CoREAS. This result is in accordance with the fact that the step sizes in ZHAireS have already been tuned from the original Aires air-shower simulation code to be sufficiently small for the Fraunhofer approximation used in the ZHS algorithm. 

A thorough analysis of the influence of the step sizes in ZHAireS is desirable, but beyond the scope of this paper. However, we can use the result of Fig.~\ref{fig:coreas_stepfc} to estimate by how much the radiation energy in ZHAireS can potentially increase with decreasing step size. The largest difference is found at small altitudes and amounts to a factor $1.4$ larger step sizes in ZHAireS than in CORSIKA with STEPFC = 0.05. We can interpret this as an effective STEPFC parameter of $0.05 \times 1.4 = 0.07$. Interpolating between the data points of Fig.~\ref{fig:coreas_stepfc} and reading off the value for STEPFC = 0.07, we find that the radiation energy could be $\sim$1\% higher if smaller step sizes were used in ZHAireS. For higher altitudes, the potential increase becomes even smaller. Hence, we can conclude that the differences in step sizes in CoREAS (with STEPFC = 0.05) and ZHAireS do not lead to differences larger than 1\%.

In the following, we will carry out the comparison of the two codes using the default setting for the simulation step size in ZHAireS,
but the optimized STEPFC = 0.05 setting for CoREAS. This ensures an adequately small step size in both codes and allows for a direct comparison.

\section{CoREAS-ZHAireS Comparison}
The radiation is almost solely emitted by the electromagnetic part of the air shower. Due to the
much lower charge-to-mass ratio, muons hardly emit radiation and can be neglected. Hence, the emitted
radiation correlates best with the energy of the electromagnetic cascade.
A measurement of the invisible energy, i.e., the amount of cosmic-ray energy not entering the electromagnetic cascade, as a function of the electromagnetic energy is presented in
\cite{ThePierreAuger:2013eja}. Combining both approaches allows us to compute the primary cosmic-ray energy from the energy radiated in the form of radio waves by the electromagnetic cascade.

However, the radiation energy does not depend solely on the electromagnetic shower energy but shows second-order dependences on the angle between the shower axis and the geomagnetic field which modulates the dominant geomagnetic emission, the air density at the shower maximum, and the relative strength of the charge-excess emission which is again a function of the air density at the shower maximum \cite{Glaser:2016qso}. The corrected radiation energy $S_{\text{rad}}$ can then be expressed as:
\begin{equation}
\label{eq:correctedRadiation}
\begin{split}
 S_{\text{rad}} = &\frac{E_{\text{rad}}}{a(\rho(X_{\text{max}}))^2 +
(1-a(\rho(X_{\text{max}}))^2)\sin^2\alpha} \\ \cdot
&\frac{1}{(1-p_0+p_0\exp{[p_1(\rho(X_{\text{max}})-\rho(\langle X_{\text{max}} \rangle))]})^2} \, ,
\end{split}
\end{equation}
where $\alpha$ is the geomagnetic angle, i.e. the angle between geomagnetic field and shower direction, $\rho(X_\mathrm{max})$ is the air density at the shower maximum $X_\mathrm{max}$ and $a$ denotes the relative charge-excess strength which is defined as
\begin{equation}
    a = \sin\alpha \sqrt{E_{\text{rad}}^{\text{ce}}/E_{\text{rad}}^{\text{geo}}}.
\end{equation}
Here, $E_{\text{rad}}^{\text{ce}}$ and $E_{\text{rad}}^{\text{geo}}$ are the radiation energies
originating from the charge-excess and geomagnetic components, respectively. The square root is 
taken for consistency with previous work where the electric field amplitude was used instead of the
radiation energy.

We will use the power law
\begin{equation}
\label{eq:PowerLaw}
 S_{\text{rad}} = A \cdot \SI{e7}{\eV} (E_{\text{em}}/\SI{e18}{\eV})^B
\end{equation}
to describe the dependency between the corrected radiation energy $S_{\text{rad}}$ and the electromagnetic shower energy $E_{\text{em}}$. Then, $A$ determines the absolute energy scale and the difference in $A$ of 
the fits for CoREAS and ZHAireS can be used as an estimate for the systematic uncertainty in their
predictions.

\subsection{Charge Excess Fraction}
Because we use a simple exponential scaling of the refractivity with height, we do not use the parametrization of $a(\rho)$ of \cite{Glaser:2016qso} but redo the parametrization.
The calculation of the radiation energy from the $\vec{v}\times (\vec{v}\times \vec{B})$ axis allows us to decompose the radiation energy into a geomagnetic and a charge excess part. This is because the polarization of the two emission mechanisms is orthogonal along this axis. 

For each simulated shower, the radiation energy originating from the geomagnetic and the
charge excess emission is calculated independently and the charge excess fraction is computed. It depends on
the air density at the shower maximum  $\rho(X_{\text{max}})$ near which most radiation is 
emitted \cite{Glaser:2016qso}.
Using the US standard atmosphere after Linsley, we can calculate $\rho(X_{\text{max}})$ from
$X_{\text{max}}$ and the zenith angle $\theta$.

An exponential function of the form
\begin{equation}
\label{eq:chargeexcessfraction}
 a(\rho(X_{\text{max}})) = q_0 + q_1 \cdot \exp{\left( q_2(\rho(X_{\text{max}}) - \rho(\langle
X_{\text{max}} \rangle)) \right)}
\end{equation}
is used to parametrize this dependency. Here, $\rho(\langle X_{\text{max}} \rangle)  =
\SI{0.65}{\kilo\gram\per\meter\squared}$ is the air density at the shower maximum for an average
zenith angle of \SI{45}{\degree} and an average $\langle X_{\text{max}} \rangle =
\SI{669}{\gram\per\centi\meter\cubed}$ as predicted by QGSJETII-04 for a shower energy of
\SI{1}{\EeV} and a \SI{50}{\percent} proton/\SI{50}{\percent} iron composition
\cite{1475-7516-2013-07-050}. The value is used here for consistency with previous studies \cite{Glaser:2016qso};
the origin from a different interaction model does not influence the validity of
our results.

\begin{figure}
\begin{subfigure}[c]{\linewidth}
\centering
\includegraphics[width=.9\linewidth]{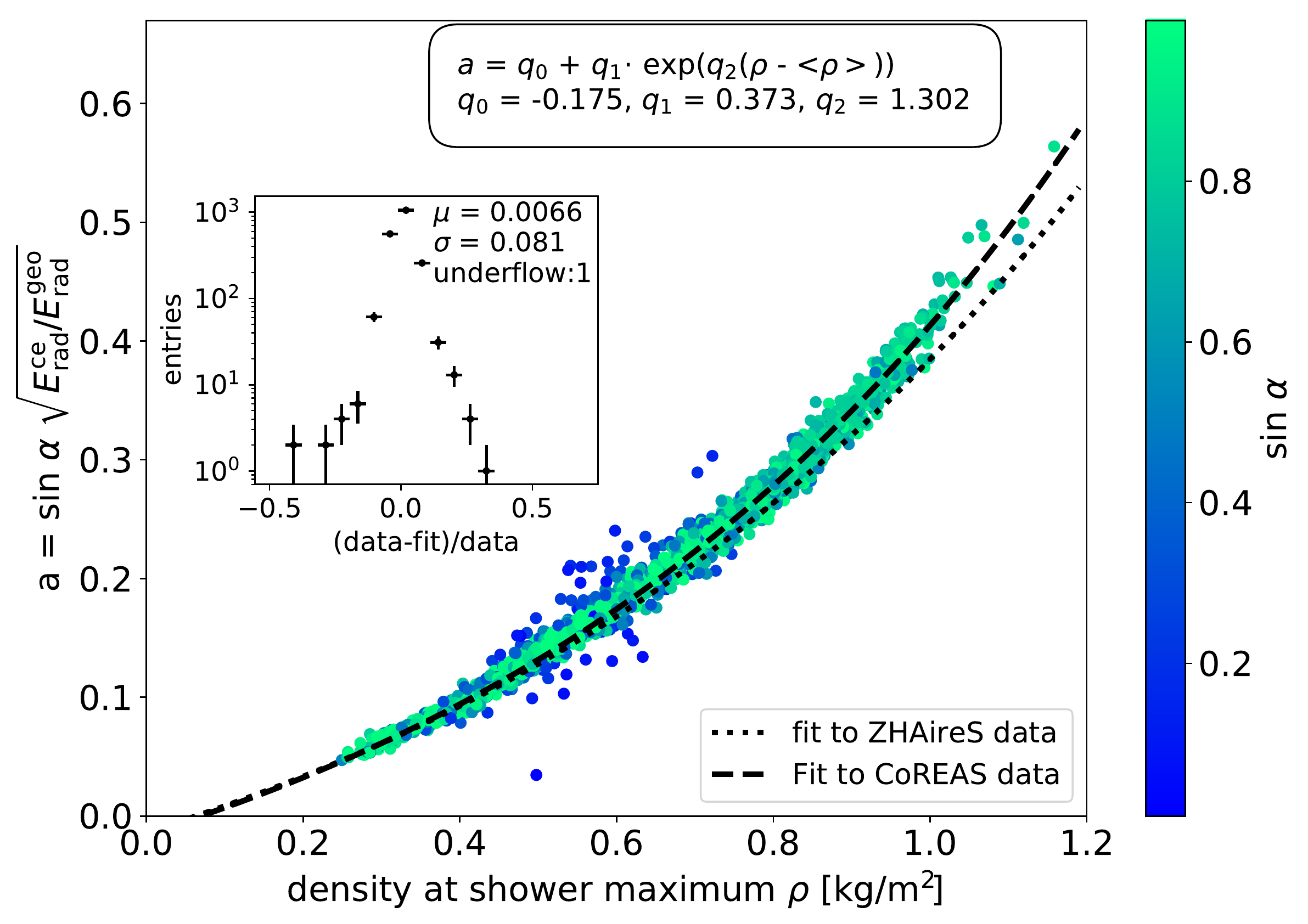}
\end{subfigure}
\begin{subfigure}[c]{\linewidth}
\centering
\includegraphics[width=.9\linewidth]{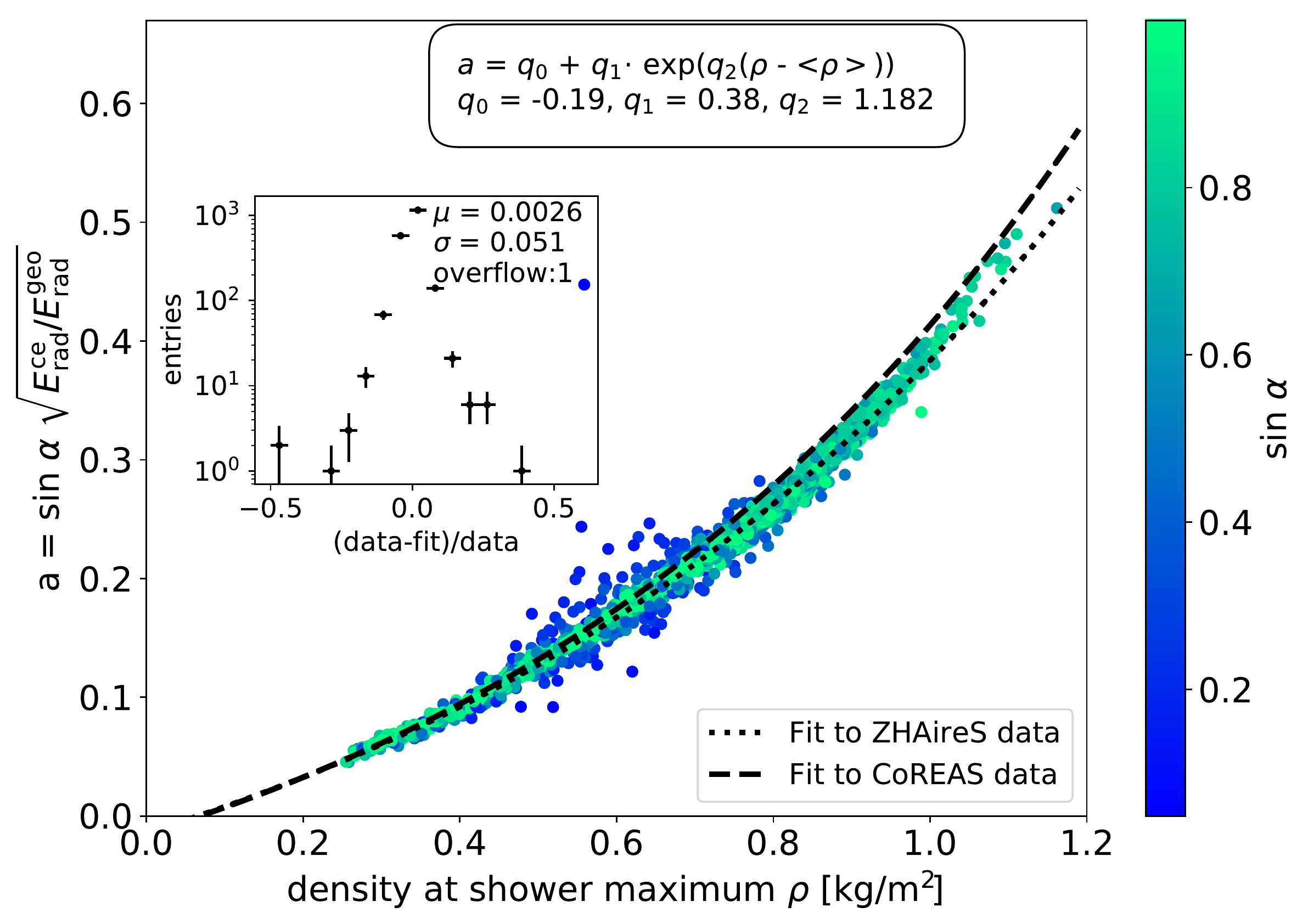}
\end{subfigure}
\caption{Charge excess fraction $a$ of air showers depending on the atmospheric density at the
shower maximum for CoREAS (top) and ZHAireS (bottom). The fit of the other code is added in both
plots for a direct comparison. The deviation between data and fit is shown in the inset figure.}
\label{fig:a}
\end{figure}

We fit Eq. \eqref{eq:chargeexcessfraction} to our data. In Fig. \ref{fig:a}, the predicted charge 
excess fractions and the fit are shown for both codes. The color code indicates that most outliers 
are air showers with a small $\sin\alpha$ value which have a limited relevance for experimental 
data.
Comparing the two fitted functions shows a good agreement between CoREAS and ZHAireS. A deviation is
found for showers with a high density at the shower maximum which corresponds to showers with small
zenith angles. The spread is \SI{8}{\percent} in the case of CoREAS simulations and \SI{5}{\percent}
for ZHAireS simulations.

\subsection{Radiation Energy}

\begin{figure}
\begin{subfigure}[c]{\linewidth}
\centering
\includegraphics[width=.9\linewidth]{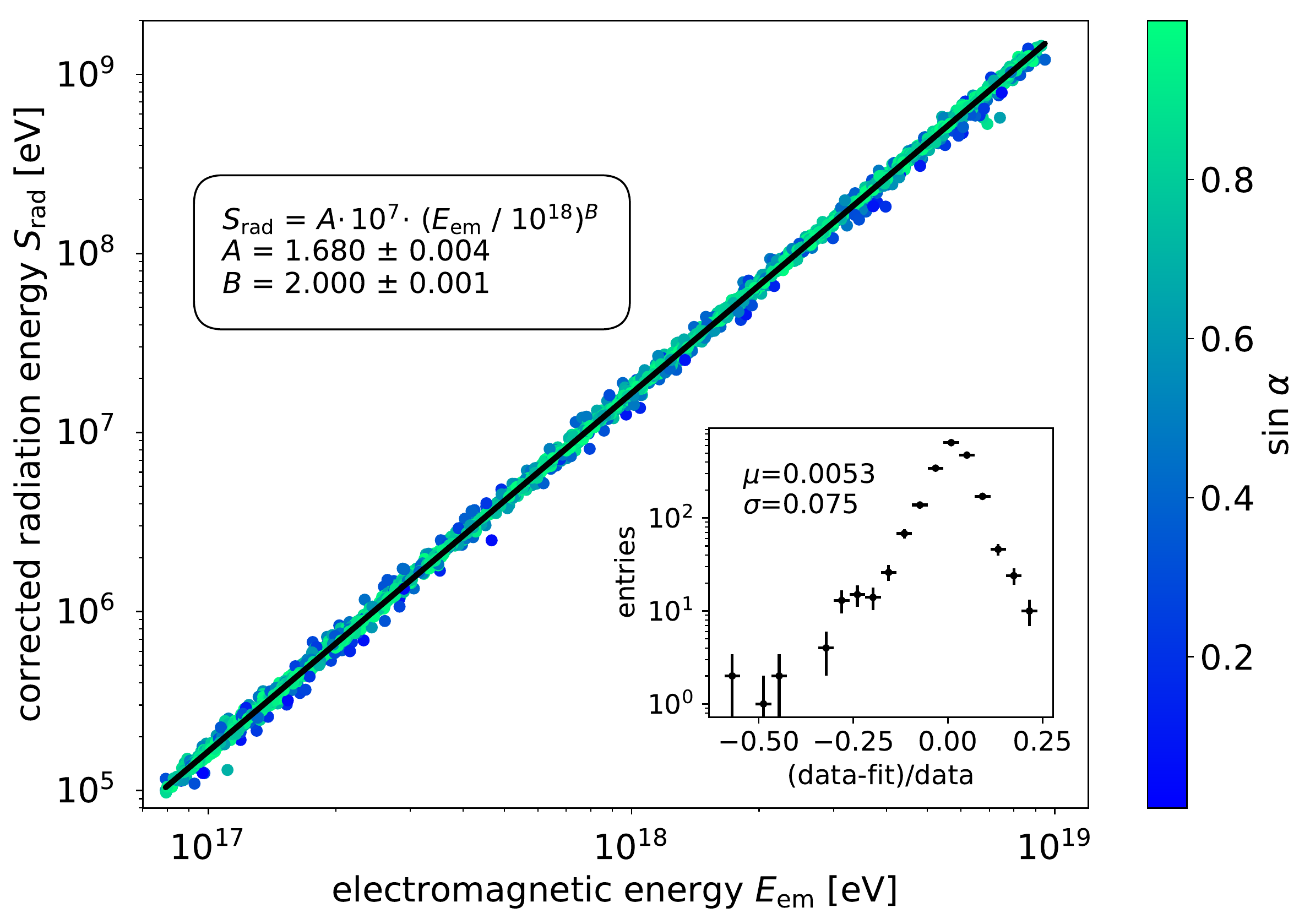}
\end{subfigure}
\begin{subfigure}[c]{\linewidth}
\centering
\includegraphics[width=.9\linewidth]{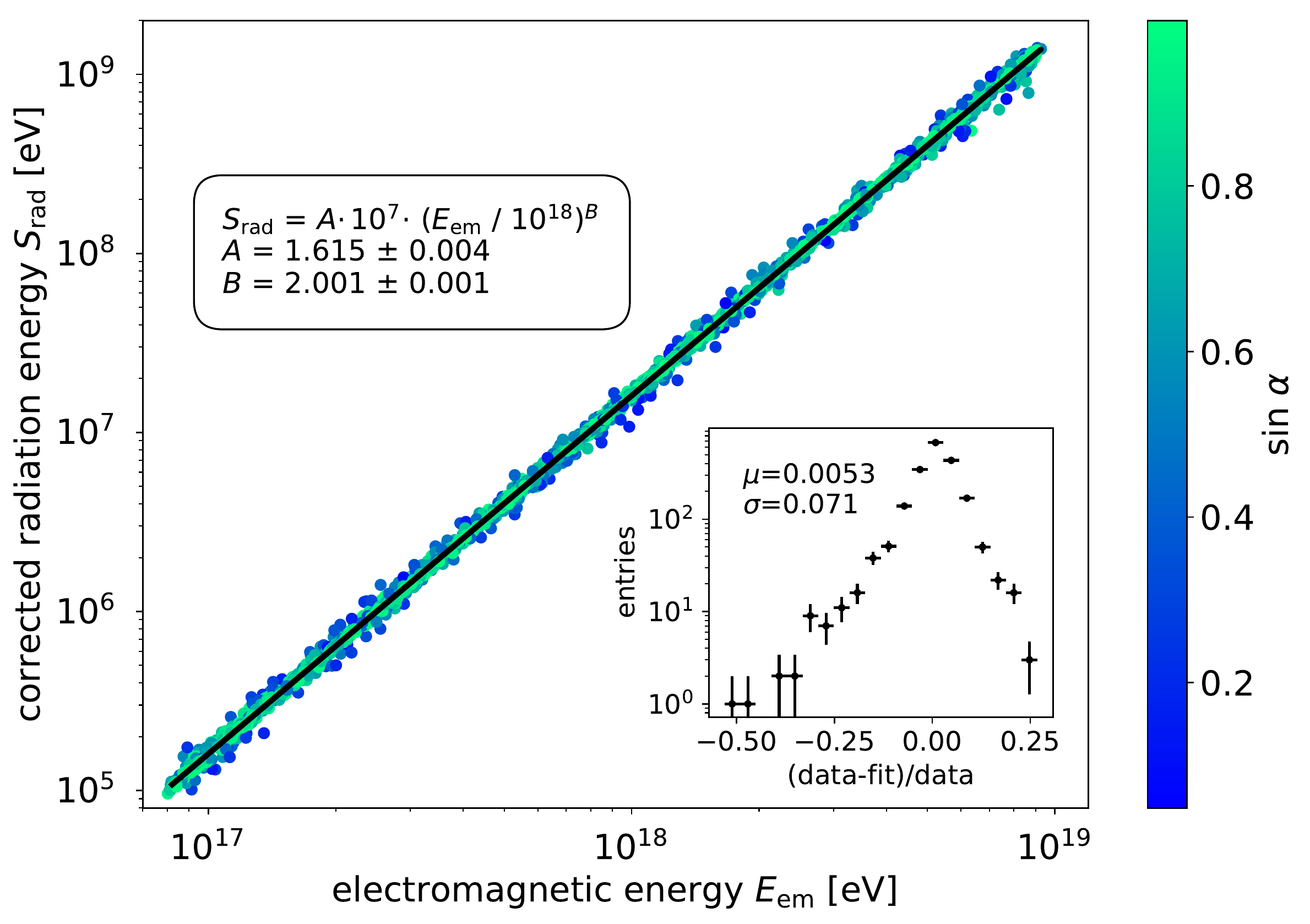}
\end{subfigure}
\caption{Correlation between the corrected radiation energy and the electromagnetic component of an
air shower for CoREAS (top) and ZHAireS (bottom).}
\label{fig:S}
\end{figure}

With the parametrization of the charge-excess strength $a$ we can use Eq.~\eqref{eq:correctedRadiation} to calculate the corrected radiation energy. For each code we use the individual fit results for the charge excess fraction. By inserting 
Eq.~\eqref{eq:correctedRadiation} in Eq.~\eqref{eq:PowerLaw} we can fit $p_0$ and $p_1$ (the parameters that describe the dependence on the air density) together with $A$ and $B$.
The results are given in table \ref{tab:fitparam}, the correlations between the corrected radiation
energy and the electromagnetic energy are shown in Fig.~\ref{fig:S}. The slope $B$ is equal
to two as it is expected for coherent emission. Looking at the deviation of data and fit a scatter
of \SI{7}{\percent} is found.
The difference in the absolute predictions of the radiation energy between CoREAS and ZHAireS is given by the ratio of 
$A_\text{CoREAS}/A_\text{ZHAireS}$. We observe a difference of \SI{4}{\percent} which corresponds 
to a difference in the electromagnetic energy of \SI{2}{\percent}.

\begin{table}
\centering
\begin{tabular}{r
   S[table-format=-1.3,table-figures-uncertainty=1]
   S[table-format=-1.3,table-figures-uncertainty=1]}
 \toprule
   & {CoREAS} & {ZHAireS}\\
 \midrule
 $A$ & 1.680(4) & 1.615(4) \\
 $B$ & 2.000(1) & 2.001(1) \\
 $p_0$ & 0.264(6) & 0.291(7) \\
 $p_1$ & -2.899(59) & -2.739(60) \\
 \bottomrule
\end{tabular}
\caption{Best fit parameters for CoREAS and ZHAireS using 1000 proton- and 1000 iron-induced
showers.}
\label{tab:fitparam}
\end{table}

\begin{table}
\centering
\begin{tabular}{r
   S[table-format=-1.4,table-figures-uncertainty=1]
   S[table-format=-1.4,table-figures-uncertainty=1]}
 \toprule
   & {CoREAS} & {ZHAireS}\\
 \midrule
 $A$ & 1.6897(11) & 1.6062(11) \\
 $B$ & 2.0012(5) & 1.9997(5) \\
 \bottomrule
\end{tabular}
\caption{Best fit parameters for CoREAS and ZHAireS with the combined corrections using 1000 proton-
and 1000 iron-induced air showers.}
\label{tab:fitparam_comb}
\end{table}

To exclude any influence on the CoREAS-ZHAireS difference from a different parametrization of $a(\rho)$ and the air-density correction (parameters $p_0$ and $p_1$), we repeat the fit of the parameters $A$ and $B$ with the same corrections for CoREAS and ZHAireS. To do so, we first repeat the fit of Eq. \eqref{eq:chargeexcessfraction} on a combined set of the charge 
excess fractions of both codes to obtain a single set of parameters $q_i$. Calculating the corrected radiation energy (Eq.~\eqref{eq:correctedRadiation}) with these parameters, we repeat the fit of the air-density correction and the power law on the 
combined data set to obtain the combined correction factors $p_i$. Then, we fit the power law 
individually for each code with the combined parameters $p_i$. Applying the combined corrections, the 
differences between CoREAS and ZHAireS are not artificially reduced by corrections optimized for
each code individually. We observe a \SI{5.2}{\percent} larger value for $A$ obtained with CoREAS in comparison with ZHAireS, which corresponds to a difference in the electromagnetic energy of \SI{2.6}{\percent}. Details of the fit are given in table 
\ref{tab:fitparam_comb}. The individual and the combined parameters are summarized in table 
\ref{tab:combined}.

\begin{table}
\centering
\begin{tabular}{r
   S[table-format=-1.3,table-figures-uncertainty=1]
   S[table-format=-1.3,table-figures-uncertainty=1]
   S[table-format=-1.3,table-figures-uncertainty=1]}
 \toprule
   & {CoREAS} & {ZHAireS} & {combined}\\
 \midrule
    $q_0$ & -0.175(98) & -0.190(118) & -0.185(77)\\
    $q_1$ & 0.373(101) & 0.380(122) & 0.379(80)\\
    $q_2$ & 1.302(376) & 1.182(399) & 1.229(241)\\
    $p_0$ & 0.264(6)& 0.291(7) & 0.280(5)\\
    $p_1$ & -2.899(59) & -2.739(60) & -2.806(44)\\
 \bottomrule
\end{tabular}
\caption{Individual and combined parameters for the corrections of the radiation energy. The $q_i$
values belong to the charge excess fraction, $p_i$ are used in the second correction term.}
\label{tab:combined}
\end{table}

\subsection{Additional Checks}
Additional checks were performed to validate the agreement between the predictions of the two
codes in more detail. We compared simulations of both
codes with different refractivities at sea level and with different magnetic field strengths. 
No significant differences between CoREAS and ZHAireS were observed.
We also checked the influence of different primary particles. Both programs predict around
\SI{3}{\percent} more radiation energy for a proton-induced air shower than for an iron-induced one with the same 
electromagnetic energy. The reason for this effect is unknown, but this interesting result implies a
sensitivity of the radio emission to the mass composition in addition to the well-known difference in the fraction of
the energy going into the electromagnetic cascade.

A last check is performed directly on the full two dimensional lateral distribution function. In
the shower plane, the LDF of the energy fluence can be described as \cite{Aab:2015vta,NellesLDF}
\begin{equation}
\label{eq:LDF2}
 f(\vec{r}) = A \left( \exp\left(-\frac{\vec{r}+C_1\vec{e}_{\vec{v} \times
\vec{B}}-\vec{r}_{\text{core}}}{\sigma^2}\right) - C_0 \exp\left(-\frac{\vec{r}+C_2\vec{e}_{\vec{v}
\times \vec{B}}-\vec{r}_{\text{core}}}{(C_3\exp(C_4\cdot\sigma))^2}\right) \right) ,
\end{equation}
where $\vec{r}$ denotes the station position, $\vec{r}_{\text{core}}$ the position of the core in
the shower plane and constants $C_i$ that are obtained by simulations and depend on the zenith
angle.
The comparison of CoREAS and ZHAireS have shown consistent results for the mean values per zenith
bin for each constant. Due to the different description of the atmosphere and the simulated
detector height we cannot compare our results with the values published in \cite{Aab:2015vta}.

\section{Conclusion}

The measurement of ultra-high energy cosmic rays via radio emission from extensive air showers allows for an accurate determination of the cosmic-ray energy scale. 
A crucial ingredient to this determination is the calculation of the radio emission from air showers on the basis of classical electrodynamics.

In this work, we have studied the accuracy of first-principle calculations of the radio emission from extensive air showers. As the radiation energy in the radio signal of air showers is a universal estimator of the cosmic-ray energy and thus independent of a specific experiment, also our estimate of the accociated systematic uncertainties is a universal quantity that applies equally to all air-shower radio experiments. Experimental efforts for the determination of the cosmic-ray energy scale on the basis of radio-emission measurements can thus directly build on our result and need not re-investigate the systematic uncertainties arising from the underlying first-principle calculations.

We determine this uncertainty by comparing the results of the two independent simulation codes CoREAS and ZHAireS. 
Technical parameters relevant to the simulations were set to the same values or were validated to not have any influence on the calculation of the radiation energy. We find that the step size chosen in the air-shower simulation has relevant influence on the predicted radio emission. For an improved estimation of the energy fluence by CoREAS, the STEPFC parameter governing the step sizes in the simulation should be reduced by a factor of $20$ to \num{0.05}. The decrease in step size results in an increase of the calculated energy fluences by 11\%. Previous comparisons between measured data and CoREAS simulations might benefit from taking this modification into account. The predicted electric field amplitudes are expected to increase by approximately \SI{5}{\percent} with
respect to the default configuration. However, such a change is well below current experimental systematic uncertainties. To estimate the influence of the simulation step size in the ZHAireS code, where it cannot be user-configured, we compared the adopted step sizes between CoREAS and ZHAireS. We found that the step sizes in ZHAireS are already close to the optimal step sizes of CoREAS so that the ZHAireS predictions of the energy fluences should not increase by more than 1\% with further decreased step sizes.

A set of 1000 proton- and 1000 iron-induced air showers was simulated to study the differences between the predictions of CoREAS and ZHAireS.
To perform the comparison of the predicted radiation energies, we corrected them for the geomagnetic angle, the relative charge-excess strength, and the air density at the shower maximum. 
The relative charge-excess strength itself depends on the air density at the shower maximum and we verified that the two simulation codes show essentially the same behavior. 
After these corrections, the radiation energy correlates well with the electromagnetic energy of the air shower. The correlation can be expressed as a quadratic power law.

We find good agreement between the absolute predictions of CoREAS and ZHAireS, with CoREAS predicting 5.2\% higher radiation energies. This comparison was made with optimized step sizes in the CoREAS code. For ZHAireS a further optimization of simulation step sizes can only decrease the observed difference by an estimated 1\%, leading to an improved agreement.
We thus conservatively quantify the difference in the radiation energy predictions of CoREAS and ZHAireS as \SI{5.2}{\percent}. 

If this deviation is interpreted as a systematic uncertainty on the absolute prediction of the radiation energy with first-principle calculations, the uncertainty in the determination of the absolute scale of the electromagnetic energy amounts to \SI{2.6}{\percent}. Monte Carlo predictions of the radiation energy of extensive air showers in the 30--80~MHz band can thus be used to set an accurate absolute energy scale for cosmic ray detectors.

\section*{Acknowledgements}

We would like to thank Mat\'ias Tueros and Jaime Alvarez-Mu\~niz for their advice and help with investigating the possibility of changing the step sizes in ZHAireS and Dieter Heck for helpful discussions regarding the step sizes adopted in CORSIKA. We furthermore acknowledge fruitful discussions with our colleagues in the AERA task of the Pierre Auger Collaboration.
Further financial support by the BMBF Verbundforschung Astroteilchenphysik and the Deutsche Forschungsgemeinschaft (DFG), grant GL 914/1-1, is acknowledged. 




\begin{thebibliography}{10}

\bibitem{Bluemer2009293}
J.~{Blümer}, R.~{Engel} and J.~R. {Hörandel}, \emph{{Cosmic rays from the
  knee to the highest energies}},
  \href{https://doi.org/10.1016/j.ppnp.2009.05.002}{\emph{Progress in Particle
  and Nuclear Physics} {\bfseries 63} (2009) 293 -- 338}.

\bibitem{EnergyScaleICRC2013}
{V. Verzi for the Pierre Auger Collaboration}, \emph{{The energy scale of the
  Pierre Auger Observatory}}, {\emph{Proc. 33rd ICRC, Rio de Janeiro, Brazil}
  (2013) }.

\bibitem{Huege:2016veh}
T.~Huege, \emph{{Radio detection of cosmic ray air showers in the digital
  era}}, \href{https://doi.org/10.1016/j.physrep.2016.02.001}{\emph{Phys.
  Rept.} {\bfseries 620} (2016) 1--52},
  [\href{https://arxiv.org/abs/1601.07426}{{\ttfamily 1601.07426}}].

\bibitem{SchroederReview}
F.~G. Schr\"oder, \emph{Radio detection of cosmic-ray air showers and
  high-energy neutrinos},
  \href{https://doi.org//10.1016/j.ppnp.2016.12.002}{\emph{Progress in Particle
  and Nuclear Physics} {\bfseries 93} (2017) 1 -- 68}.

\bibitem{Aab:2016eeq}
{\scshape Pierre Auger} collaboration, A.~Aab et~al., \emph{{Measurement of the
  Radiation Energy in the Radio Signal of Extensive Air Showers as a Universal
  Estimator of Cosmic-Ray Energy}},
  \href{https://doi.org/10.1103/PhysRevLett.116.241101}{\emph{Phys. Rev. Lett.}
  {\bfseries 116} (2016) 241101},
  [\href{https://arxiv.org/abs/1605.02564}{{\ttfamily 1605.02564}}].

\bibitem{TunkaRexCrossCalibration}
P.~A. {Bezyazeekov}, N.~M. {Budnev}, O.~A. {Gress} and {et al.}, \emph{{Radio
  measurements of the energy and the depth of the shower maximum of cosmic-ray
  air showers by Tunka-Rex}}, {\emph{Journal of Cosmology and Astroparticle
  Physics} {\bfseries 2016} (2016) 052}.
  
\bibitem{AERAOctocopterJINST}
{\scshape Pierre Auger} collaboration, A.~Aab et~al., \emph{{Calibration of the 
  logarithmic-periodic dipole antenna (LPDA) radio stations at the 
  Pierre Auger Observatory using an octocopter}},
  \href{https://doi.org/10.1088/1748-0221/12/10/T10005}{\emph{J. Instrum. }
  {\bfseries 12} (2017) T10005-T10005}.


\bibitem{KrauseICRC2017}
{R. Krause for the Pierre Auger Collaboration}, \emph{A new method to determine
  the energy scale for high-energy cosmic rays using mhz radio measurements at
  the pierre auger observatory},  in \emph{Proceedings of the 35th ICRC, Busan,
  Korea}, 2017.

\bibitem{coreas}
T.~Huege, M.~Ludwig and C.~James, \emph{Simulating radio emission from air
  showers with coreas}, {\emph{AIP Conf. Proc.} (2013) 128--132}.

\bibitem{zhaires}
J.~Alvarez-Muñiz, W.~R. Carvalho, Jr. and E.~Zas, \emph{Monte carlo
  simulations of radio pulses in atmospheric showers using {ZHAireS}},
  \href{https://doi.org/10.1016/j.astropartphys.2011.10.005}{\emph{Astroparticle
  Physics} {\bfseries 35} (2012) 325 -- 341}.

\bibitem{Aab:2015vta}
{\scshape Pierre Auger} collaboration, A.~Aab et~al., \emph{{Energy Estimation
  of Cosmic Rays with the Engineering Radio Array of the Pierre Auger
  Observatory}}, \href{https://doi.org/10.1103/PhysRevD.93.122005}{\emph{Phys.
  Rev.} {\bfseries D93} (2016) 122005},
  [\href{https://arxiv.org/abs/1508.04267}{{\ttfamily 1508.04267}}].

\bibitem{Glaser:2016qso}
C.~Glaser, M.~Erdmann, J.~R. Hörandel, T.~Huege and J.~Schulz,
  \emph{{Simulation of Radiation Energy Release in Air Showers}},
  \href{https://doi.org/10.1088/1475-7516/2016/09/024}{\emph{JCAP} {\bfseries
  1609} (2016) 024}, [\href{https://arxiv.org/abs/1606.01641}{{\ttfamily
  1606.01641}}].

\bibitem{2011PhRvE..84e6602J}
C.~W. {James}, H.~{Falcke}, T.~{Huege} and M.~{Ludwig}, \emph{{General
  description of electromagnetic radiation processes based on instantaneous
  charge acceleration in ``endpoints''}},
  \href{https://doi.org/10.1103/PhysRevE.84.056602}{\emph{Physical Review E}
  {\bfseries 84} (Nov., 2011) 056602},
  [\href{https://arxiv.org/abs/1007.4146}{{\ttfamily 1007.4146}}].

\bibitem{Zas:1991jv}
E.~Zas, F.~Halzen and T.~Stanev, \emph{{Electromagnetic pulses from high-energy
  showers: Implications for neutrino detection}},
  \href{https://doi.org/10.1103/PhysRevD.45.362}{\emph{Phys. Rev.} {\bfseries
  D45} (1992) 362--376}.

\bibitem{2010PhRvD..81l3009A}
J.~{Alvarez-Mu{\~n}iz}, A.~{Romero-Wolf} and E.~{Zas}, \emph{{{\v C}erenkov
  radio pulses from electromagnetic showers in the time domain}},
  \href{https://doi.org/10.1103/PhysRevD.81.123009}{\emph{Physical Review D}
  {\bfseries 81} (June, 2010) 123009},
  [\href{https://arxiv.org/abs/1002.3873}{{\ttfamily 1002.3873}}].

\bibitem{SLACT510-PRL}
{\scshape T-510} collaboration, K.~Belov, K.~Mulrey, A.~Romero-Wolf, S.~A.
  Wissel, A.~Zilles, K.~Bechtol et~al., \emph{Accelerator measurements of
  magnetically induced radio emission from particle cascades with applications
  to cosmic-ray air showers},
  \href{https://doi.org/10.1103/PhysRevLett.116.141103}{\emph{Phys. Rev. Lett.}
  {\bfseries 116} (Apr, 2016) 141103}.

\bibitem{corsika}
D.~Heck, J.~Knapp, J.~N. Capdevielle, G.~Schatz and T.~Thouw, \emph{{CORSIKA: A
  Monte Carlo Code to Simulate Extensive Air Showers}}, .

\bibitem{Schroder:2016imf}
{F. Schröder for the Pierre Auger Collaboration}, \emph{{Radio detection of
  high-energy cosmic rays with the Auger Engineering Radio Array}},
  \href{https://doi.org/10.1016/j.nima.2015.08.047}{\emph{Nucl. Instrum. Meth.}
  {\bfseries A824} (2016) 648--651},
  [\href{https://arxiv.org/abs/1601.00462}{{\ttfamily 1601.00462}}].

\bibitem{ThePierreAugerCollaboration2015172}
{The Pierre Auger Collaboration}, \emph{{The Pierre Auger Cosmic Ray
  Observatory}}, \href{https://doi.org/10.1016/j.nima.2015.06.058}{\emph{Nucl.
  Instrum. Meth.} {\bfseries A798} (2015) 172 -- 213}.

\bibitem{Alvarez-Muniz:2014wna}
J.~Alvarez-Muñiz, W.~R. Carvalho, Jr., H.~Schoorlemmer and E.~Zas,
  \emph{{Radio pulses from ultra-high energy atmospheric showers as the
  superposition of Askaryan and geomagnetic mechanisms}},
  \href{https://doi.org/10.1016/j.astropartphys.2014.04.004}{\emph{Astropart.
  Phys.} {\bfseries 59} (2014) 29--38},
  [\href{https://arxiv.org/abs/1402.3504}{{\ttfamily 1402.3504}}].

\bibitem{Kobal2001}
M.~{Kobal}, \emph{{A thinning method using weight limitation for air-shower
  simulations}},
  \href{https://doi.org/10.1016/S0927-6505(00)00158-4}{\emph{Astropart. Phys.}
  {\bfseries 15} (June, 2001) 259--273}.

\bibitem{fluka}
G.~Battistoni, F.~Cerutti, A.~Fass{\`{o}}, A.~Ferrari, S.~Muraro and J.~R.
  et~al., \emph{The fluka code: description and benchmarking}, {\emph{AIP Conf.
  Proc.} (2007) 31--49}.

\bibitem{0954-3899-25-9-308}
M.~Bleicher, E.~Zabrodin, C.~Spieles, S.~A. Bass, C.~Ernst, S.~Soff et~al.,
  \emph{Relativistic hadron-hadron collisions in the ultra-relativistic quantum
  molecular dynamics model}, {\emph{Journal of Physics G: Nuclear and Particle
  Physics} {\bfseries 25} (1999) 1859}.

\bibitem{MunizTuerosPrivateComms}
M.~{Tueros} and J.~{Alvarez-Mu\~niz}. private communication.

\bibitem{ThePierreAuger:2013eja}
{M. Tueros for the Pierre Auger Collaboration}, \emph{{Estimate of the
  non-calorimetric energy of showers observed with the fluorescence and surface
  detectors of the Pierre Auger Observatory}},  in \emph{{Proceedings, 33rd
  International Cosmic Ray Conference (ICRC2013): Rio de Janeiro, Brazil, July
  2-9, 2013}}, pp.~11--14, 2013,
  \href{https://arxiv.org/abs/1307.5059}{{\ttfamily 1307.5059}},
  \href{http://lss.fnal.gov/archive/2013/conf/fermilab-conf-13-285-ad-ae-cd-td.pdf}
  {http://lss.fnal.gov/archive/2013/conf/fermilab-conf-13-285-ad-ae-cd-td.pdf}.

\bibitem{1475-7516-2013-07-050}
M.~D. Domenico, M.~Settimo, S.~Riggi and E.~Bertin, \emph{Reinterpreting the
  development of extensive air showers initiated by nuclei and photons},
  {\emph{Journal of Cosmology and Astroparticle Physics} {\bfseries 2013}
  (2013) 050}.

\bibitem{NellesLDF}
A.~Nelles, S.~Buitink, H.~Falcke and {et al.}, \emph{{A parameterization for
  the radio emission of air showers as predicted by CoREAS simulations and
  applied to LOFAR measurements}},
  \href{https://doi.org/10.1016/j.astropartphys.2014.05.001}{\emph{Astropart.
  Phys.} {\bfseries 60} (2015) 13 -- 24}.

\end{thebibliography}


\section*{References}

\providecommand{\href}[2]{#2}\begingroup\raggedright\endgroup

\end{document}